\documentclass{mem}
\usepackage{natbib}
\usepackage{txfonts}
\usepackage{balance}
\usepackage{graphicx}
\usepackage[a4paper]{hyperref}
\idline{75}{282}
\begin{document}
\def\teff{$T\rm_{eff }$}
\def\kms{$\mathrm {km s}^{-1}$}

\title{Improving Large-scale Convection Zone-to-Corona Models}

\subtitle{}

\author{W.~P. \, Abbett\inst{1} \and G.~H. \, Fisher\inst{1}}

\offprints{W.~P. Abbett}

\institute{Space Sciences Laboratory, University of California, 
Berkeley, CA 94720-7450 \email{abbett@ssl.berkeley.edu}}

\authorrunning{Abbett \& Fisher}

\titlerunning{Improving Large-scale Models}

\abstract{We introduce two new methods that are designed to improve the 
realism and utility of large, active region-scale 3D MHD models of the solar 
atmosphere.  We apply these methods to RADMHD, a code capable of modeling 
the Sun's upper convection zone, photosphere, chromosphere, transition 
region, and corona within a single computational volume.  We first present 
a way to approximate the physics of optically-thick radiative transfer without
having to take the computationally expensive step of solving the radiative 
transfer equation in detail.  We then briefly describe a rudimentary 
assimilative technique that allows a time series of vector magnetograms to 
be directly incorporated into the MHD system. 
}
\maketitle{}

\section{Introduction}
In this paper, we briefly summarize our efforts to improve our models of quiet
Sun and active region magnetic fields in computational domains that include the 
upper convection zone, photosphere, chromosphere, transition region and low corona 
within a single computational domain. Our goal is similar to that presented in 
\citet{Abbett07} --- that is, to develop the techniques necessary to efficiently 
simulate the spatially and temporally disparate convection zone-to-corona interface 
over spatial scales sufficiently large to accommodate at least one active region.

The advantage of this type of single-domain modeling is clear. For example, 
evolving a turbulent convection zone and corona simultaneously in a physically 
self-consistent way allows for the quantitative study of important physical 
processes such as flux emergence, submergence and cancellation; the transport of 
magnetic free energy and helicity into the solar atmosphere; the generation of 
magnetic fields via a convective dynamo; and the physics of coronal heating. 

However, this approach is challenging.  The computational domain is highly
stratified --- average thermodynamic quantities change by many orders of 
magnitude as the domain transitions from a relatively cool, turbulent
regime below the visible surface, to a hot, magnetically-dominated and 
shock-dominated regime high in the model atmosphere.  In addition, the low 
atmosphere is where the radiation field transitions from being optically thick 
to optically thin.  The chromosphere itself presents an additional 
challenge, since the radiation field is often decoupled from the thermal pool, 
particularly in some of the strongest, most energetically important transitions.

There are a number of ways to model the energetics
of the convection zone-to-corona system, ranging from approximate, parameterized 
descriptions of the thermodynamics (see e.g., \citealt{Fan09,Hood09}), to highly 
realistic treatments of radiative transfer (see e.g., \citealt{Martinez09a,
Martinez09b}). Since our objective is to model the coupled system over large 
spatial scales, our goal is to find the most efficient treatment of the 
energetics possible that still provides a physically meaningful representation 
of the dynamic connection between the convection zone and corona.  

In order to describe the thermodynamics of the corona, a 
model should include the effects of electron thermal conduction along
magnetic field lines and radiative cooling in the optically-thin limit.  
In addition, some physics-based or empirically-based source of coronal heating 
must be present if the model corona is to remain hot.  In the 
convective interior well below the visible surface, radiative cooling can be
treated in the diffusion limit. The trick is, how best to describe
the effects of optically-thick radiative transfer in the region of the model
atmosphere that lies between these two extremes.

The most satisfying approach would be to couple the LTE transfer equation (or 
non-LTE population and transfer equations) to the MHD system to obtain cooling rates
and intensities that could be compared directly to observations. Unfortunately,
for large active region or global-scale problems, the computational expense of
these techniques remains prohibitive.

In \citet{Abbett07} we tried the opposite approach --- ignore the transfer 
equation altogether, and develop an artificial, fully parameterized means of 
approximating surface cooling (in this case, we employed a modified form of Newton 
cooling).  This worked relatively well, provided we carefully calibrated the 
adjustable parameters to match the average sub-surface stratification of previous, 
more realistic simulations of magneto-convection where the LTE transfer equation 
was solved in detail \citep{Bercik02}.

Of course, the principle drawback of this approach 
is that it is ultimately \emph{ad hoc} and unphysical, and requires 
other, more realistic simulations as a basis for calibration in order to get 
meaningful results.  We therefore have developed an approximation that
is based on the macroscopic radiative transfer equation, and have incorporated 
this new treatment into our 3D MHD model, RADMHD.  We describe this new
method in Section 2.

While it is important to treat the energetics of the system in a physically 
meaningful way, it is also important to remember that the utility of a given 
simulation ultimately depends on the statement of the problem.  For an MHD 
simulation, this boils down to one's choice of initial states and boundary 
conditions.  It is of great benefit, for example, to pose a simple, well-defined 
problem, and set up a numerical experiment that can shed light on what is 
believed to be the relevant physical processes in an otherwise complex system.  
For example, important progress has been made in understanding the physics of 
magnetic flux emergence by studying how idealized twisted flux ropes emerge 
through highly-stratified model atmospheres (see e.g., \citealt{Cheung07,Fan04}). 

Yet the observed evolution of the photospheric magnetic field is often
far more complex, particularly in and around CME and flare producing active 
regions.  It is very difficult to set up a simple magnetic and energetic
configuration that can initialize a simulation that will faithfully 
mimic the observed evolution of a real active region.  It is desirable to do so, 
however, since we wish to quantitatively understand the physical mechanisms of
energy storage and release, and the transport of magnetic energy and helicity 
between the convective interior and corona.   

To make progress, we could take a cue from meteorologists, and investigate a 
means to incorporate observational data directly into MHD models.  This 
is not at all straightforward for solar models however, since data is obtained 
entirely through remote sensing, and not \emph{in situ}.

To address this challenge, we have developed a simple, rudimentary means 
of assimilating a time series of vector magnetograms into an MHD model of
the photosphere-to-corona system.  We briefly summarize this technique in
Section 3, and apply it to the specific problem of finding a 3D magnetic 
field that is as force-free as possible given a single measurement of the 
vector magnetic field at the photosphere. 

\section{An Approximate Treatment of Optically Thick Cooling}

What follows is a brief description of our approximate treatment of 
optically-thick radiative cooling in the portion of the computational domain that 
represents the solar photosphere and chromosphere. In practice, this cooling is 
incorporated into the MHD system as a source term in the equation that evolves the 
internal energy per unit volume (Equation 4 of \citealt{Abbett07}).  

We begin by characterizing the net cooling rate for a volume of plasma at some 
location within the solar atmosphere:

\begin{equation}
  R=\! \int \! d\Omega \! \int \! d\nu \left(\eta_\nu-\kappa_\nu I_\nu\right) .
\end{equation}

Here, $\nu$ represents frequency, and $\Omega$ solid angle. The emissivity, 
opacity, and specific intensity are frequency dependent, and are denoted 
$\eta_\nu$, $\kappa_\nu$ , and $I_\nu$ respectively. Rearranging the order of 
integration, and defining the source function $S_\nu$ as the ratio of the 
emissivity to opacity, we have

\begin{figure}[]
\resizebox{\hsize}{!}{\includegraphics[clip=true]{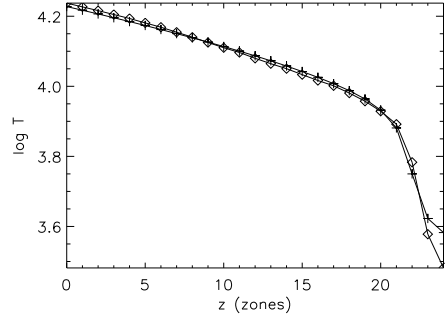}}
\caption{
\footnotesize Shown is a comparison of the average temperature stratification
between the realistic radiative magneto-convection simulations of \citet{Bercik02}
(\emph{crosses}), and the RADMHD model convection zone using the new treatment of
optically-thick transfer (\emph{diamonds}).
}
\label{figure1}
\end{figure}

\begin{equation}
  R= \! \int \! d\nu \, \kappa_\nu  \! \int \! d\Omega \left(S_\nu-I_\nu\right) .
\end{equation}

\noindent Since the source function is independent of direction, we recast the 
integral as

\begin{equation}
  R=4\pi  \! \int \! d\nu \, \kappa_\nu \left( S_\nu-J_\nu \right)
\end{equation}

\noindent with mean intensity

\begin{equation}
  J_\nu \equiv \frac{1}{4\pi}\int \! d\Omega \, I_\nu .
\end{equation}

\noindent The formal solution for the specific intensity in the plane-parallel 
approximation is

\begin{equation}
  I_\nu (\mu)=\int_{0}^{\infty} d\tau^{\prime} \frac{e^{-|\tau_\nu
    -\tau^{\prime}|/|\mu|}}{|\mu |}S_\nu(\tau^{\prime}) ,
\end{equation}

\noindent where $\mu$ is the usual cosine angle. Then the mean intensity can 
be expressed as

\begin{equation}
  J_\nu =\frac{1}{2}\int_{0}^{\infty} d\tau^{\prime}S_\nu (\tau^{\prime}) 
    \int_0^1 d\,|\mu |\frac{e^{-|\tau_\nu-\tau^{\prime}|/|\mu|}}{|\mu |} .
\end{equation}

\noindent The integral over $\mu$ can now be evaluated and the mean intensity 
can be cast as

\begin{equation}
  J_\nu =\frac{1}{2}\int_{0}^{\infty} d\tau^{\prime}S_\nu (\tau^{\prime})
    E_1(|\tau_\nu-\tau^{\prime}|) ,
\end{equation}

\begin{figure*}[t!]
\resizebox{\hsize}{!}{\includegraphics[clip=true]{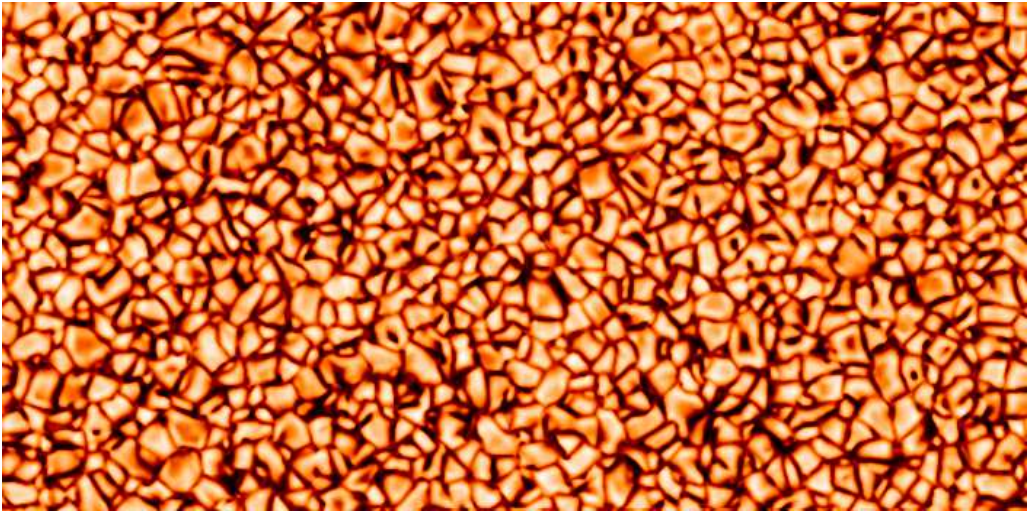}}
\caption{\footnotesize The temperature along a horizontal slice through a 
RADMHD model photosphere during the relaxation process. This simulation uses 
the \emph{ad hoc} Newton cooling described in \citet{Abbett07} to approximate 
optically-thick surface cooling.  The slice spans $75\times 37.5$ Mm$\,^2$.}
\label{figure2}
\end{figure*}

\noindent where $E_1$ denotes the first exponential integral. So far, this 
is simply textbook radiative transfer (e.g., \citealt{Mihalas78}). No 
approximations have yet to be made, other than an assumption of a locally 
plane-parallel atmosphere. Now we'll make our first approximation. Note that 
$E_1(|\tau_\nu-\tau^{\prime}|)$ is singular when $\tau^{\prime}=\tau_\nu$, 
and that the singularity is integrable. Since $E_1$ is peaked around
$\tau_\nu$, contributions from $S_\nu(\tau^\prime)$ will be centered 
around  $S_\nu(\tau_\nu)$. Thus, we approximate the mean intensity by

\begin{equation}
  J_\nu \approx \frac{1}{2}S_\nu(\tau_\nu)\int_{0}^{\infty}d\tau^{\prime} 
    E_1(|\tau_\nu-\tau^{\prime}|) .
\end{equation}

This integral can then be evaluated, giving a simple expression for the 
mean intensity,

\begin{equation}
  J_\nu \approx S_\nu(\tau_\nu)\left( 1-\frac{E_2(\tau_\nu)}{2}\right) ,
\end{equation}

\noindent where $E_2$ refers to the second exponential integral. Note 
that this can be rewritten in the following way: 

\begin{equation}\label{eqn10}
  1-\frac{J_\nu}{S_\nu} \approx \frac{E_2(\tau_\nu)}{2} .
\end{equation}

\noindent We now return to our expression for the net cooling rate and
recast it in a slightly different form,

\begin{equation}
  R=4\pi \! \int \! d\nu \, \kappa_\nu S_\nu
    \left(1-\frac{J_\nu}{S_\nu}\right) .
\end{equation}

\noindent Substituting equation~\ref{eqn10} into the integrand, we have

\begin{equation}
  R\approx 2\pi \! \int \! d\nu \, \kappa_\nu S_\nu E_2(\tau_\nu) .
\end{equation}

\noindent If we further assume LTE, the source function is simply the 
Planck function and we have

\begin{equation}
  R\approx 2\pi \! \int  \! d\nu \, \kappa_\nu B_\nu (T) E_2(\tau_\nu) .
\end{equation}

\begin{figure*}[t!]
\resizebox{\hsize}{!}{\includegraphics[clip=true]{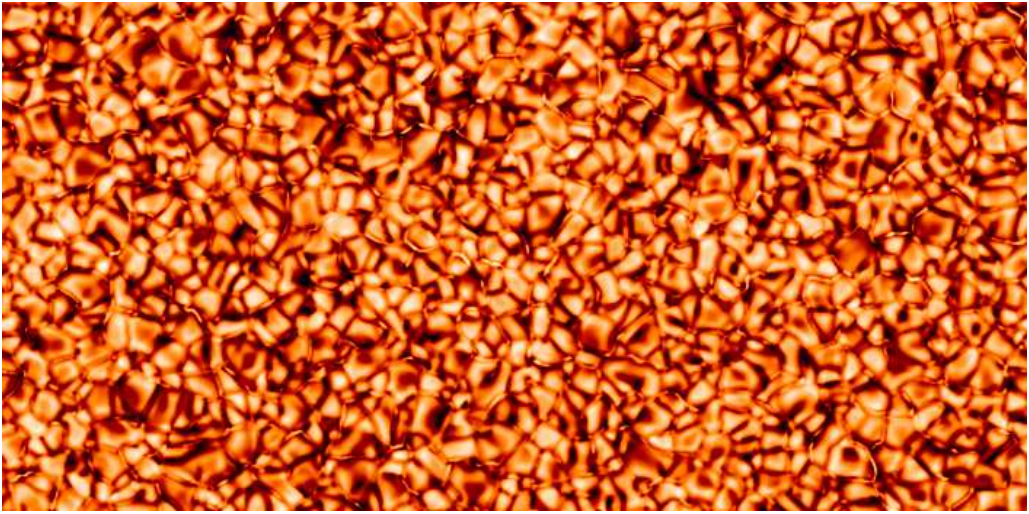}}
\caption{\footnotesize The temperature along a horizontal slice through a 
RADMHD model photosphere during the relaxation process.  In this case, we
use our new treatment to approximate the optically-thick surface cooling.
We note that the model convection zone has yet to fully relax.}
\label{figure3}
\end{figure*}

Now for the real swindle! Let's integrate over frequency, and replace the
frequency-dependent opacity by its Planck weighted mean value: That is,
replace $\kappa_\nu B_\nu$ with $\overline{\kappa}(\sigma/\pi)T^4$ where 
$\overline{\kappa}$ represents a Planck-weighted mean opacity, and $\sigma$
the Stefan-Boltzmann constant. Then including the exponential function in 
the average over frequency, we find that

\begin{equation}
  R\approx 2C\,\overline{\kappa}\sigma T^{\,4} E_2(\alpha\overline{\tau}) .
\end{equation}

\noindent Here, $C$ represents the normalization constant for the 
integration.  The arbitrary constant $\alpha$ appears in the exponential 
integral since the mean opacity used in the calculation of the optical 
depth scale could differ in general from the mean opacity that appears 
by itself in the integrand.

To determine the normalization constant $C$, we integrate our cooling 
function from zero to infinity over an isothermal slab to obtain the 
total radiative flux. The resulting expression must be equal to the 
known result $F_{tot}=\sigma T^{\,4}$. This allows us to determine the 
normalization constant $C=\alpha$.  To evaluate $\alpha$, we compare
the detailed cooling rate depth distribution using this formulation with 
the cooling rate in the \citet{Bercik02} LTE model of the solar atmosphere, 
and conclude the best-fit value is $\alpha =1$ (See Figure~\ref{figure1}). 
Thus, our approximate cooling function takes the form:

\begin{equation}
  R\approx 2\,\overline{\kappa}\sigma T^{\,4} E_2(\overline{\tau}) .
\end{equation}

The advantage of this treatment lies in its simplicity. The above 
approximation for surface cooling, while non-linear, is trivial to calculate 
for each mesh element.  It is certainly more physical than the \emph{ad hoc}
treatment employed in \citet{Abbett07}, since it is based on the radiative 
transfer equation, and incorporates an optical depth scale into the model. 
Further, it has no adjustable parameters.  The only calibration now required 
is a choice of optical depth ranges over which to apply the different 
approximations.  Currently, we use the radiative diffusion approximation 
for optical depths greater than 10, an expression for radiative cooling in 
the optically-thin limit for optical depths less than 0.1, and the
above treatment in the intervening layers.

Figures~\ref{figure2} and \ref{figure3} provide a qualitative comparison between 
a model convection zone generated using the \emph{ad hoc} approach of 
\cite{Abbett07} to estimate the effects of optically-thick radiative 
surface cooling, and one that utilizes the approximation described above.
The images represent the gas temperature along a horizontal slice 
through a RADMHD model photosphere during the relaxation process.
Both simulations used the same initial convective state and boundary
conditions (periodic in the horizontal directions and closed vertically); 
the only difference is the treatment of the optically-thick transfer.  
Distinct differences rapidly develop --- the convective cells become more 
irregularly shaped while the size distribution of cells begins to more 
closely mimic that of the more realistic simulations of \citet{Bercik02}.  

However, there are irregularities in the current data set.  For example, 
there are regions within the intergranular lanes that are hotter than 
expected.  This may be an artifact resulting from our empirically-based 
coronal heating function (see \citealt{Abbett07} for details) extending 
unphysically deep into the atmosphere (i.e., its optical depth cutoff is 
too high), or it may simply be a transient effect that will subside as 
the simulation progresses.  This is a work in progress, and we
continue to test and validate the new treatment against our previous
quiet Sun simulations, and against more realistic magnetoconvection
simulations that treat the LTE transfer equation in detail.

\section{Rudimentary Data Assimilation}

We now turn our attention to the problem of incorporating a time
series of vector magnetic field measurements into our MHD model of
the solar atmosphere.  The essential problem is that even the most
carefully pre-processed sequences of vector magnetograms cannot be 
expected to exactly satisfy Faraday's law, and thus are physically
inconsistent from the point of view of the numerical model.  

This is not particularly surprising, since it is a non-trivial task 
to properly transform polarization measurements into a vector 
magnetic field.  The datasets naturally suffer from the effects 
of uncertainty due to noise, seeing, or saturation; the inversion process 
itself is model dependent; and the well-known 180 degree ambiguity
in the transverse field must be resolved in the context of a timeseries 
of magnetograms rather than in a single magnetogram in isolation.

\begin{figure*}[t!]
\resizebox{\hsize}{!}{\includegraphics[clip=true]{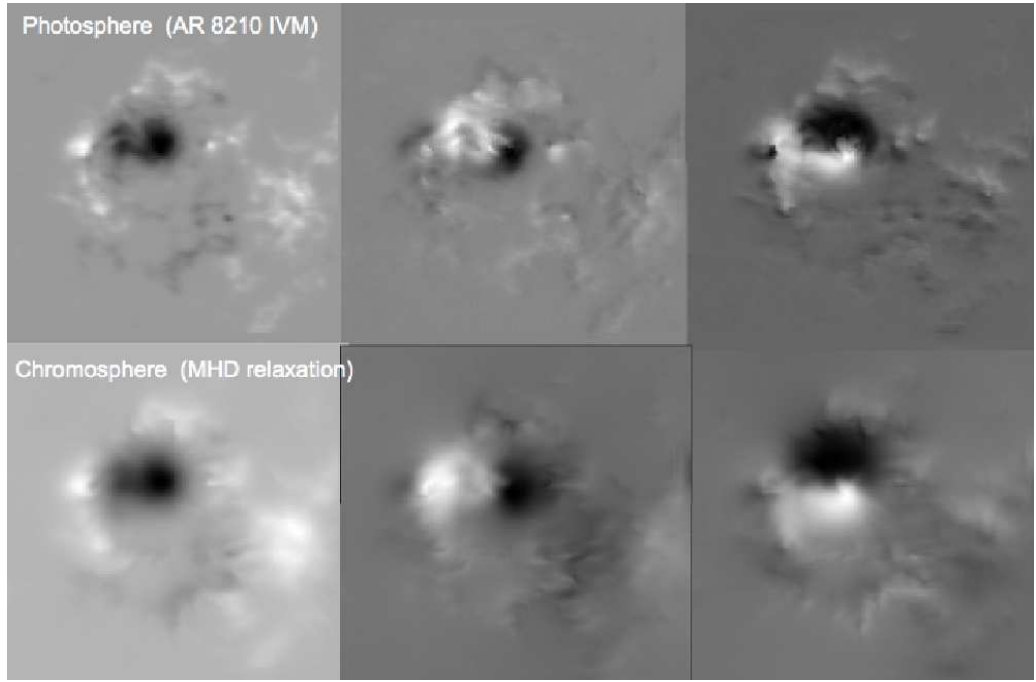}}
\caption{\footnotesize The three components of the magnetic field
(from left to right $B_z$, $B_x$, and $B_y$) from the early stages
of a 3D RADMHD simulation of NOAA AR 8210.  The top row shows a
horizontal slice through the model photosphere, and the bottom row
shows the magnetic field through a horizontal slice corresponding
to the model chromosphere.  The top row mirrors the vector magnetic 
field as measured by the IVM vector magnetograph at Mees Solar 
Observatory on Haleakala, and the bottom row represents a synthetic 
chromospheric magnetogram as predicted by the MHD model.}
\label{figure4}
\end{figure*}

While the data itself presents challenges, it is important to remember
that the model suffers from fundamental deficiencies of its own.  The
single-fluid MHD system may not capture the essential physics of the solar 
atmosphere, particularly in the low atmosphere where effects such as 
non-LTE transfer, ion-neutral diffusion, magnetic reconnection, and 
non-thermal partical acceleration may play fundamental roles in the 
dynamic evolution of a given region.  

Then how shall we proceed?  One approach is to take the 
data at face value and incorporate the measurements directly into 
an MHD model in the form of time-dependent, characteristic boundary 
conditions \citep{Wu06}.  Here, one must be mindful to not over-specify the 
MHD system.  This method is restrictive in that only certain components 
of the electric field or flow inferred from the data can be used to drive
the simulation.  In addition, one must make assumptions about the 
thermodynamics of the system in order to drive the model atmosphere
in a physical way.  

Here, we take another approach.  To avoid the 
mathematical constraints inherent to MHD boundary conditions, we push 
our lower boundary slightly deeper into the photosphere and instead 
incorporate the data into the model via additional forces acting on 
active zones of the calculation where the entire MHD system is 
being self-consistently evolved.  

To do this, we must first obtain an inductive flow field from a given 
time series of magnetograms that is both consistent with the observed 
evolution of the vector field and Faraday's law.  This is a non-trivial 
task, as the problem is inherently under-determined, and the cadence and 
quality of the magnetograms may vary.  There is a growing number of inversion 
techniques that address this problem \citep{Fisher09,Ravindra08,Schuck08,
Georgoulis06,Welsch04,Longcope04,Kusano02}; each is capable of providing
an inductive flowfield consistent with the observations and suitable for 
incorporation into an MHD model.  

Next, we must generate an initial near-equilibrium or steady state
atmosphere from the first magnetogram of the timeseries, and choose an
appropriate set of exterior boundary conditions.  The challenge here is
to minimize perpendicular currents within the computational volume, and
to provide coronal boundary conditions that minimize forces resulting from
magnetic tension.  This way, the forces introduced into the model 
photosphere are the principle drivers of the system.  

As a starting point, we 
generate a non-constant-$\alpha$ force-free extrapolation using a variation 
of the optimization technique of \citet{Wheatland00}.  Given the
photospheric magnetogram and a choice of external boundary conditions,
this procedure minimizes the functional

\begin{equation}\label{opt}
  f=\int dV \; \left( \frac{|({\bf\nabla\times B})\,{\bf \times \, B}|^2}
    {B^2}+|{\bf\nabla\cdot B}|^2 \right) ,
\end{equation}

\noindent and generates an initial magnetic configuration.  

Unfortunately, the reality is that the photosphere is often far from force-free, 
making the mathematical problem of generating perfectly force-free equilibia 
ill-posed.  While the optimization technique performs well relative to other
methods (see \citealt{Schrijver06}), it still cannot be expected 
to fully converge to an equilibrium state without altering
the transverse magnetic field at the photospheric boundary.

We therefore use the optimization method to generate an initial starting 
point for an MHD relaxation.  The above functional need not
be vanishingly small in every mesh element, since the MHD code will
diffuse away any significant divergence error, and clean up any noisy, 
unphysical currents near the lower boundary (see Figure~\ref{figure4}).  In 
practice, this is done by artificially damping fast-moving waves and allowing 
the system to slowly evolve to a near-equilibrium state.  
Of course, the resulting atmosphere is not expected to be force-free near 
the photosphere.  The currents in the system are, however, more physical
since they were evolved via the MHD system of conservation equations 
rather than by attempting to minimize the functional of equation~\ref{opt}.

Our purpose here is not to find a perfectly stable equilibrium solution.
In fact, such a state may not exist, given the vector magnetogram and 
choice of boundary conditions.  We are simply striving for an
initial atmosphere that is not so vastly out of force balance that motions at
the model photosphere are immediately overwhelmed by other less relevant 
processes.  Once this is achieved, we drive the atmosphere in the following 
way.

First, we define the physical contribution to the force as that described 
by the MHD momentum conservation equation (see \citealt{Abbett07} for 
details),

\begin{equation}
{\bf F}\equiv -{\bf\nabla\cdot}\left[\rho
   {\bf uu}\!+\!\left(p+\frac{B^2}{8\pi}\right){\bf I}\!-\!\frac{{\bf BB}}
   {4\pi}\!-\!{\bf\Pi}\right]\!+\!\rho{\bf g} 
\end{equation}

\noindent We then define the forces implied by the data,

\begin{equation}
{\bf F}_{data}\equiv \frac{\partial\rho{\bf u}_{\mathrm{inv}}}
   {\partial t} .
\end{equation}

\noindent Here, ${\bf u}_{\mathrm{inv}}$ refers to the inductive flow field 
obtained through one of the many velocity inversion techniques (see 
\citealt{Welsch07}).  

Then in a thin volume corresponding to the model's photosphere, we recast 
the momentum equation in the following form:

\begin{equation}
\frac{\partial\rho{\bf u}}{\partial t}{\Big\vert}_{phot}=\xi
   ({\bf F}_{data})_\perp+\left(1-\xi\right)({\bf F})_\perp
   +({\bf F})_{||} 
\end{equation}

\noindent where the parallel and perpendicular subscripts denote the forces 
parallel or perpendicular to the direction of the magnetic field.
Here, $0<\xi<1$ represents a ``confidence matrix'' defined at each mesh
element within the photospheric volume.  It is easy to see that when 
$\xi=1$, the forces perpendicular to the magnetic field within the
model photosphere are determined entirely by the data, and when $\xi=0$, 
the photospheric layer evolves as the MHD system normally would in 
the absence of any observational forcing.  

Since flows parallel to the field do not affect magnetic evolution, we 
allow them to evolve in an unconstrained fashion.  All other independent 
variables evolve as prescribed by the MHD system of equations 
\emph{including} the magnetic field. Recall, ${\bf u}_{\mathrm{inv}}$
is designed to be consistent with of the observed evolution of 
one or more components of the photospheric magnetic field (depending
on the inversion method used) and, in principle, should drive the photospheric 
field in a manner consistent with the timeseries of magnetograms.

\section{Concluding Remarks}

We have developed a rudimentary means of assimilating a 
time series of vector magnetograms into the interior volume of an MHD 
model in a manner that is stable, and does not over-specify the problem.
We are currently using this assimilative technique to incorporate
a timeseries of vector magnetograms into a 120 Mm$\,^3$ RADMHD 
model atmosphere that contains a model photosphere, chromosphere, 
transition region and corona.  The IVM data we are using is a four
hour timeseries from NOAA AR 8210 --- a well-studied flare 
and CME-producing active region.  The simulations are in their 
preliminary stages, and we hope to report on this work in the
near future.

In addition, we have presented a computationally 
efficient method of approximating optically-thick radiative cooling 
in our RADMHD quiet Sun models.   The treatment improves upon the 
method of \citet{Abbett07}, while still retaining the efficiency 
necessary to allow for large, active region-scale, convection 
zone-to-corona computational domains.  Our simulations are progressing, 
and we are currently evaluating the efficacy and reliability of the 
new method.  We are optimistic that each of these methods will 
improve the realism and utility of our current suite of numerical models. 

\begin{acknowledgements}
The authors would like to thank Brian Welsch and K.D. Leka for
providing us with the timeseries of vector magnetograms of NOAA 8210.
This ongoing work is supported in part by the NASA TR\&T and 
Heliophysics Theory program, and by the National Science Foundation 
through the SHINE and ATM programs.  Many of the simulations described 
here were performed on NASA’s NCCS Discover supercomputer.
\end{acknowledgements}

\bibliographystyle{aa}

\end{document}